\documentclass[showpacs,prl,aps,superscriptaddress,twocolumn,floatfix]{revtex4}
\usepackage{graphicx,float,amsmath,amssymb}
\newfont{\ensmathquatorze}{msbm10 scaled 1400}
\newfont{\ensmathonze}{msbm10 scaled 1100}
\newfont{\ensmathdix}{msbm10}
\newfont{\ensmathneuf}{msbm10 scaled 833}
\newfont{\ensmathhuit}{msbm10 scaled 694}
\newfam\ensmathfam                        
\textfont\ensmathfam=\ensmathonze        
\scriptfont\ensmathfam=\ensmathdix       
\scriptscriptfont\ensmathfam=\ensmathhuit
\def\sss{\scriptscriptstyle}
\newcommand{\APPROX}[1]{                % approximativement egal a .. si ..
   {{\raisebox{-.3cm}{$\textstyle\simeq$}} \atop {\scriptstyle{#1}}}}

\begin{document}

\title{Propagation of a Dark Soliton in a Disordered Bose-Einstein Condensate}

\author{Nicolas Bilas}
\author{Nicolas Pavloff}
\affiliation{Laboratoire de Physique Th\'eorique
et Mod\`eles Statistiques,
Universit\'e Paris Sud, b\^at. 100, F-91405 Orsay Cedex, France}

\begin{abstract}
We consider the propagation of a dark soliton in a quasi 1D
Bose-Einstein condensate in presence of a random potential. This
configuration involves nonlinear effects and disorder, and we argue that,
contrarily to the study of stationary transmission coefficients through a
nonlinear disordered slab, it is a well defined problem. It is found that a
dark soliton decays algebraically, over a characteristic length which is
independent of its initial velocity, and much larger than both the healing
length and the 1D scattering length of the system. We also determine the
characteristic decay time.
\end{abstract}

%\noindent03.75.-b Matter waves\hfill\break
%\noindent 05.60.Gg Quantum transport\hfill\break
%\noindent 42.65.Tg Optical solitons; nonlinear guided waves\hfill\break

\maketitle

%%%%%%%%%%%%%%%%%%%%%%%%%%%%%%%%%%%%%%%%%%%%%%%%%%%%%%%%%%%%%%%%%%%%%%%%%%%%%

Phase coherent systems display wave mechanical pro\-per\-ties distinct from those
typically observed at macroscopic scale. In particular, transport in presence
of disorder is strongly affected by interference effects, leading to weak or
strong localization, as observed in many different fields (electronic or
atomic physics, acoustics, electromagnetism). Our understanding of these
effects have made great progresses over the last decades in the case of
non-interacting linear waves. Some studies have consi\-de\-red the propagation of
plane waves or bright solitons in a disordered region in the case of
attractive interaction (see, e.g., the review \cite{Gre92} and the discussion
below of the results of Ref.~\cite{Kiv90}), but almost nothing is known in the
case of repulsive nonlinearity.

The field of Bose condensed atomic vapors allows new investigations of such
phenomena in presence of repulsive or attractive interaction, in an
intrinsically phase coherent system over which the experimental control is
rapidly progressing. Such studies have begun with the observation of
``fragmentation of the condensate'' over a microchip \cite{corrug}; random
potentials have recently been engineered using an optical speckle pattern
\cite{Lye04}; and it has also been proposed to implement disorder by using two
different atomic species in an optical lattice \cite{gav04}.

In the present Letter, we study the transport properties of a quasi one
dimensional (1D) Bose-Einstein condensate in pre\-sen\-ce of disorder and
repulsive interaction. The configuration we study corresponds to the ``1D mean
field regime'' \cite{Men02}, where the system is described by a 1D
order para\-me\-ter $\psi(x,t)$ depending on a single spatial variable: the
coordinate $x$ along the direction of propagation. $\psi(x,t)$ obeys the non
linear Schr\"odinger equation
\begin{equation}\label{e1}
i\hbar\, \frac{\partial \psi}{\partial t}
= -\frac{\hbar^2}{2m}\frac{\partial^2\psi}{\partial x^2} +
\Big(U(x)+g\,|\psi|^2\Big) \psi \; ,
\end{equation}
where $U(x)$ is the random potential and $g$ an effective coupling constant
which reads $g=2\hbar\omega_\perp a$ in the case of particles experiencing an
effective repulsive interaction characterized by the 3D s-wave scattering
length $a$ ($a>0$), and a transverse harmonic confinement with pulsation
$\omega_\perp$ \cite{Ols98}. It is customary to define the oscillator length
$a_\perp=(\hbar/m\omega_\perp)^{1/2}$ and $a_1=a_\perp^2/2a$ ($-a_1$ is the 1D
scattering length \cite{Ols98}). Denoting by $n_{1\sss D}$ a typical value of
$|\psi(x,t)|^2$, the 1D mean field regime corresponds to a situation where
$1\ll n_{1\sss D} a_1\ll (a_1/a_\perp)^2$. The first inequality ensures that
the system does not get in the Tonks-Girardeau limit and the second that the
transverse wave-function is the ground state of the linear transverse
Hamiltonian \cite{Men02,Pet04}.

A particular issue specific to Eq.~(\ref{e1}) is the very possibility to
define a transmission coefficient. Since the equation is nonlinear, it is not
possible in general to disentangle an incident and a reflected current in the
region upstream the potential (in other words, a reflected atom will interact
with the incident beam) \cite{Leb03}. A possible way for avoiding this
problem is to change the transverse confining potential upstream the
disordered potential so that, in this region, nonlinear effects become
negligible \cite{Leb03}. However, even in this case, a technical difficulty
arises because of multistability: several stationary solutions exist for a
given input state \cite{Gre92,Leb03}. Moreover, in the case of repulsive
interaction we consider here, for extended enough disordered
region, no stationary solution exist and the transmission coefficient can only
be defined via a time average \cite{Pau05}.

A way out of these difficulties consists in studying the propagation of a
soliton in the system. This constitutes an intrinsically time-dependent
problem, but the input and output states can be precisely characterized, and the
transmission is simply defined by comparing the large time behaviors
($t\to\pm\infty$) of the solution. This route has been followed by Kivshar
{\it et al.} \cite{Kiv90} in the case of {\it attractive} nonlinearity
($g<0$). In this case, a solitary wave is a bright soliton, characterized by
two parameters: its velocity $V$ and the number of particles $N$ inside the
soliton. The disordered potential was taken as
\begin{equation}\label{e2}
U(x)=g_{\mbox{\tiny{imp}}} \, \sum_n \delta (x - x_n) \; ,\quad
\mbox{where}\quad
g_{\mbox{\tiny{imp}}}=\frac{\hbar^2}{mb}\; .
\end{equation}
$U(x)$ describes a series of static impurities with equal intensity and random
positions $x_n$. The $x_n$'s are uncorrelated and uniformly distributed with
mean density $n_{\mbox{\tiny{imp}}}$. In this case $<\!\!U(x)\!\!> =
g_{\mbox{\tiny{imp}}} n_{\mbox{\tiny{imp}}}$ and $<\!\!U(x_1)U(x_2)\!\!> -
<\!\!U(x_1)\!\!>\times<\!\!U(x_2)\!\!> = (\hbar^2/m)^2 D \,\delta(x_1-x_2)$,
with $D=n_{\mbox{\tiny{imp}}}/b^2$. From what is known in the case of linear
waves, this type of potential is typical insofar as localization properties
are concerned. In the weakly nonlinear regime $mV^2/2\gg \hbar^2 N^2/(m
a_1^2)$, it was found in Ref.~\cite{Kiv90} that the soliton velocity remains
approximatively constant in the disordered region, whereas $N$ shows an
exponential decay similar to what occurs for a linear wave packet. In the
opposite strongly nonlinear regime, it was found that the soliton behaves very
differently, leading asymptotically to a configuration where both $N$ and $V$
become practically constant (independent of $x$).

In the present Letter, we consider the case of {\it repulsive} nonlinearity
where the solitary waves are dark solitons. We find that the propagation of
these solitons in a disordered potential is quite peculiar for the two
following reasons: first, the strongly and weakly nonlinear cases cannot be
considered as distinct because, in a given system, the number of particles
forming the soliton cannot evolve independently from its velocity; and
secondly, a dark soliton has a velocity bounded by the velocity of sound in
the system. As a result, dark solitons behave differently from the bright ones
studied in Ref.~\cite{Kiv90}: initially rather ``nonlinear'' solitons decay
algebraically (and not exponentially), becoming eventually ``linear''. Besides
the length covered by the soliton in the disordered region is independent of
its initial velocity.

\

\includegraphics*[width=8cm]{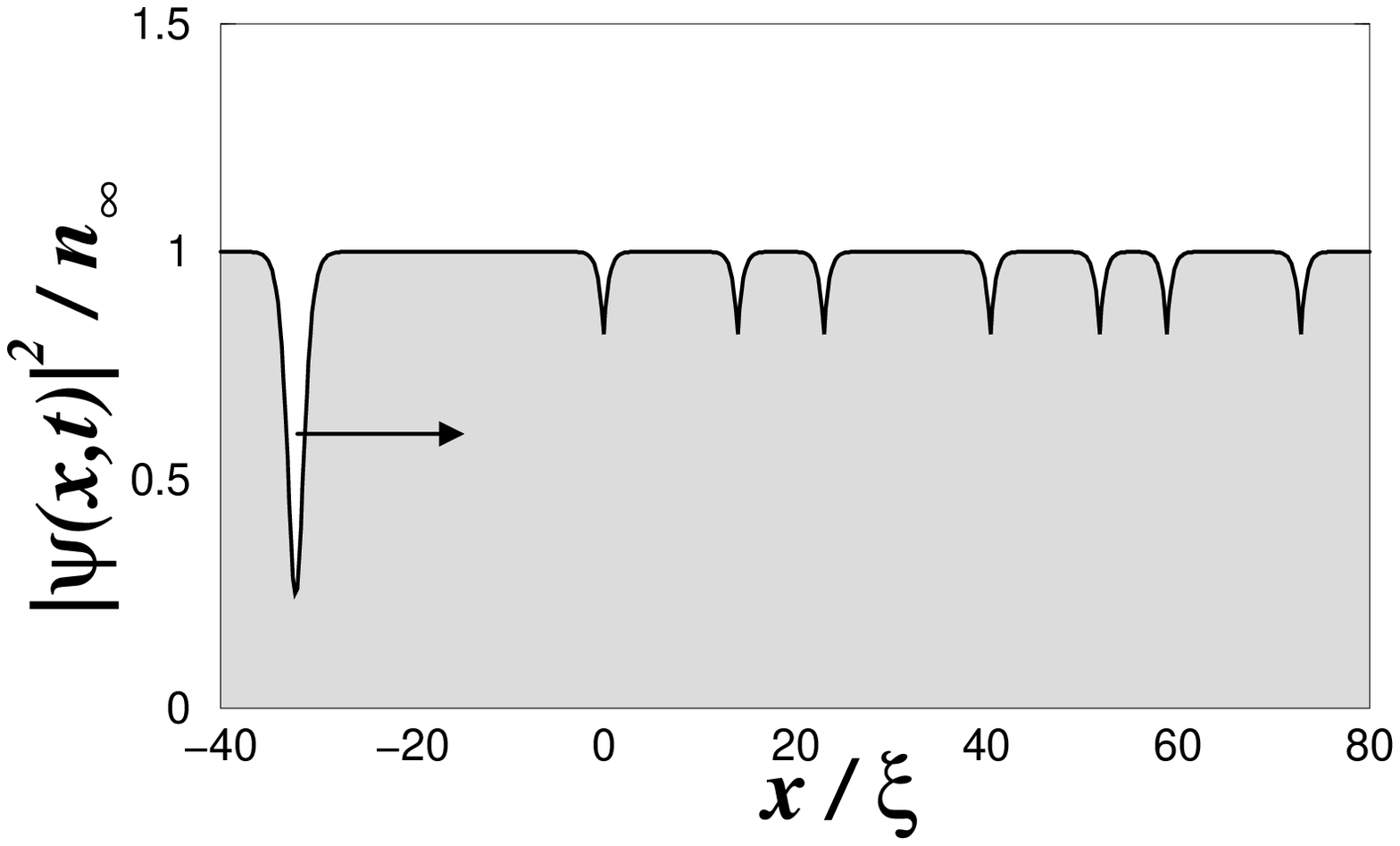}

Figure 1: {\sl Density profile of a dark soliton incident with velocity
$V=c/2$ on point-like repulsive obstacles with random positions corresponding
to a potential $U(x)$ given in Eq.~(\ref{e2}), with $\xi/b=0.2$ and
$n_{\mbox{\tiny{imp}}}\,\xi\simeq 0.1$. The arrow
represents the direction of propagation of the soliton.}

\

Let us thus consider a dark soliton with initial velocity $V$, incident from
the left on a disordered potential of type (\ref{e2}), with $x_0=0<x_1<...$.
This situation is illustrated in Fig. 1. The soliton is characterized by two
parameters, its velocity $V$ and the asymptotic background density
$n_\infty=\lim_{x\to\pm\infty}|\psi(x,t)|^2$. Instead of $n_\infty$, one can
equivalently employ the chemical potential $\mu=g\,n_\infty$, the healing
length $\xi=(a_1/n_\infty)^{1/2}$, or
the speed of sound $c = \hbar/m\xi$. A dark
soliton has a velocity $V\le c$, an energy
\begin{equation}\label{esol}
E_{\mbox{\tiny{sol}}}=\frac{4}{3}\mu \left(\frac{a_1}{\xi}\right)
\left(1-\frac{V^2}{c^2}\right)^{3/2} \; ,
\end{equation}
\noindent and consists in a density trough of typical extension
$\xi\,(1-V^2/c^2)^{-1/2}$, corresponding to a number of missing particles
$\Delta N=2\,(a_1/\xi) (1-V^2/c^2)^{1/2}$. In the
1D mean field regime where (\ref{e1}) holds, $a_1\gg\xi$ and $\Delta N$ is
typically a large number, except in the limit where $V$ is close to $c$. This
occurs at velocities around
$V_{\mbox{\tiny{crit}}}=c\,[1-(\xi/2a_1)^2]^{1/2}$. At such velocities $\Delta
N\sim 1$ and the soliton has an extension $\sim a_1$.

We consider the case where the average separation between the impurities is
much larger than the healing length ($n_{\mbox{\tiny{imp}}}\xi\ll 1$) and the
initial velocity of the soliton is not close to $c$. In this case, the
scattering of the soliton from the impurities can be treated as a sequence of
independent events.
When the soliton encounters a single obstacle, it radiates phonons which form
two counter propagating wave packets moving at velocity $c$. Accordingly,
its energy decreases by an amount $\delta E_{\mbox{\tiny{sol}}}=-
E_{\mbox{\tiny{rad}}}^+ - E_{\mbox{\tiny{rad}}}^-$, where
$E_{\mbox{\tiny{rad}}}^+$ ($E_{\mbox{\tiny{rad}}}^-$) is the forward
(backward) emitted energy. It was found in Ref.~\cite{Bil05} that
\begin{equation}\label{e3}
E_{\mbox{\tiny{rad}}}^\pm=\mu \left(\frac{\xi}{b}\right)^2
F^\pm(V/c)\; ,
\end{equation}
where $F^\pm(v)$ is a dimensionless function defined for
$v=V/c\in[0,1]$ as
\begin{equation}\label{e4}
F^\pm(v)=\frac{\pi}{16 \, v^6}\, 
\int_0^{+\infty}\!\!\!\!dy \;
\frac{y^4\left(-v\pm\sqrt{1+y^2/4}\,\right)^2}
{\sinh^2
\left(\frac{\pi\,y\,\sqrt{1+y^2/4}}
{2\,v\,\sqrt{1-v^2}}\right)
}
\; .
\end{equation}
Equation (\ref{e3}) is a perturbative result valid in the limit $b\gg\xi$ and
$V^2\gg c^2 (\xi/b)$. The first inequality ensures that the impurity only weakly
perturbs the static background and the second that the scattering of the soliton
by the impurity can be treated perturbatively.
The soliton having lost energy during the collision, its velocity changes by
an amount $\delta V=c\,\delta v$ which, from (\ref{esol}), is related to
$\delta E_{\mbox{\tiny{sol}}}$ via $v\,\delta v\, (1-v^2)^{-1}=- \frac{1}{3}
\delta E_{\mbox{\tiny{sol}}}/E_{\mbox{\tiny{sol}}}$. 

Since $n_{\mbox{\tiny{imp}}}\xi\ll 1$, one can go to the continuous limit
considering the successive collisions as a sequence of random uncorrelated
events. Over a length $\delta x$ the solitons will experience
$n_{\mbox{\tiny{imp}}} \delta x$ such collisions. This leads to the following
differential equation:
\begin{equation}\label{e7}
\frac{dv}{dx}=\frac{1}{4\,x_0}\;\frac{F^+(v)+F^-(v)}{v\,\sqrt{1-v^2}} \; ,
\end{equation}
\noindent where $x_0=a_1b^2/(\xi^3\,n_{\mbox{\tiny{imp}}})=a_1/(D\,\xi^3)$.
Equation (\ref{e7}) can be solved analytically in the high velocity regime,
when $v\to1$. In this limit, $F^+(v)+F^-(v)\simeq\frac{4}{15}(1-v^2)^{5/2}$
and, for a soliton of initial velocity $V_{\mbox{\tiny{init}}}$ one obtains
\begin{equation}\label{e8}
\frac{V(x)}{c}=\left\{
1-
\frac{1-(V_{\mbox{\tiny{init}}}/c)^2}
{1+\left[1-(V_{\mbox{\tiny{init}}}/c)^2\right]\frac{2\,x}{15\,x_0}}
\right\}^{1/2} \; .
\end{equation}
We compare in Figure 2 the results of this approximate solution with the
numerical solution of Eq.~(\ref{e7}) in the cases $V_{\mbox{\tiny{init}}}/c=$
0.75, 0.5 and 0.25. The agreement is very good, even for initial velocities
which are not close to $c$.

\

\includegraphics*[width=8cm]{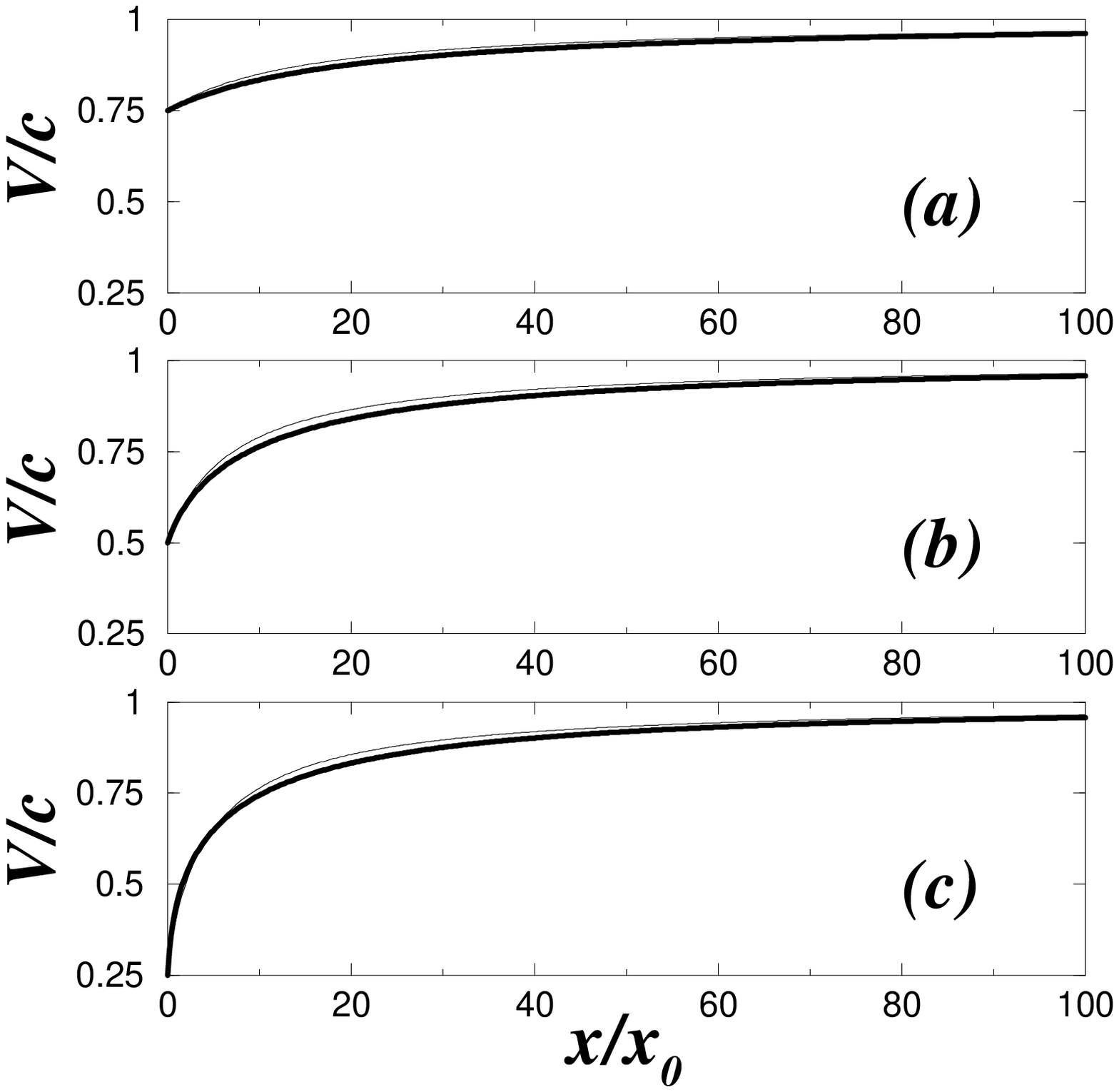}

Figure 2: {\sl Evolution of the velocity $V$ of a dark soliton as a function
of the distance $x$ traveled in the disordered region. In each plot, the thick
line corresponds to the numerical solution of Eq.~(\ref{e7}) and the thin line
to Eq.~(\ref{e8}). Case (a) corresponds to an initial velocity
$V_{\mbox{\tiny{init}}}=3\,c/4$, case (b) to $V_{\mbox{\tiny{init}}}=c/2$ and
case (c) to $V_{\mbox{\tiny{init}}}=c/4$ .}

\

The soliton is accelerated as it progresses through the disordered region (as
seen in Figure 2) because it radiates energy at each collision with an
impurity. This increased velocity after a loss of energy is a typical feature
of dark solitons which can be considered as particles with a negative kinetic
mass which decreases with increasing energy \cite{Fed99} (see
Eq.~(\ref{esol})). One also notices in Figure 2 that $V$ saturates when it
gets close to $c$, meaning that, in this regime, the rate of energy loss
decreases. The reason for this phenomenon is that a dark soliton cannot have a
velocity higher than $c$. As a result, when its velocity reaches this upper
bound, the soliton cannot lose a large fraction of its energy, because this
would lead, after the collision, to an unphysical value of $V$ (larger than
$c$). This phenomenon has an important consequence on the maximum distance $L$
over which the soliton can travel in the disordered region. As seen on Figure
2, $L$ is very large and seems independent from the initial velocity of the
soliton. In order to get a quantitative evaluation, we define $L$ as being the
length after which the soliton is a trough containing only one particle, i.e.,
the velocity $V(L)$ in Eq.~(\ref{e8}) reaches the value
$V_{\mbox{\tiny{crit}}}$. In this limit the soliton can no
longer be detected by standard imaging techniques, and for all practical
purposes one can consider that it has totally decayed. From (\ref{e8}) one
obtains
\begin{equation}\label{e9}
L=\frac{15\,x_0}{2}\left(\frac{2a_1}{\xi}\right)^2=
30\,a_1\,\frac{(a_1/\xi)^2}{D\,\xi^3}\; .
\end{equation}
This confirms what was inferred from Figure 2: a slow soliton will initially
decay more rapidly than a fast one and altogether, the distance over which
solitons can travel before completely decaying is independent of their initial
velocity. As expected, $L$ decreases for increasing disorder, the
effect of the disorder being measured by the dimensionless parameter
$D\,\xi^3$, i.e., by the two points correlation function of the random
potential. Irrespectively of the value of the parameter
$D\,\xi^3$, we remark that $L$ is large compared with
$a_1$, since in the 1D mean field regime $a_1\gg \xi$. Hence, a dark soliton
covers quite a large distance in the disordered region before decaying.

The distance $L$ is covered in a time $\tau$ which we now evaluate. It is
important to realize that $V$ is not the average velocity of the soliton, but
its velocity between two obstacles: in vicinity of an impurity, the velocity
of the soliton decreases if $g_{\mbox{\tiny{imp}}} >0$ and increases if
$g_{\mbox{\tiny{imp}}} <0$. As a result, the asymptotic position of the
soliton is shifted compared to what it would be in absence of obstacle. In the
case of a single impurity, this shift $\Delta$ can be quite accurately
evaluated by means of the ``effective potential theory'' as being \cite{Bil05}
\begin{equation}\label{e10}
\Delta=\int_{-\infty}^{+\infty}\!\! dx
\left[1-\frac{1}{\sqrt{1-U_{\mbox{\tiny eff}}(x)/mV^2}} \right]\; ,
\end{equation}
\noindent where $U_{\mbox{\tiny eff}}$ is an effective potential which
reads in the case of a point like impurity $U_{\mbox{\tiny eff}}(x)=
\frac{g_{\mbox{\tiny{imp}}}}{2\xi}\cosh^{-2}(x/\xi)$. In the limit $V^2\gg
c^2(\xi/b)$ where Eq. (\ref{e3}) holds, the shift reads $\Delta\simeq
-c^2\xi^2/(2bV^2)$. In the presence of multiple impurities, going to the
continuous limit, one obtains that during a time $\delta t$ the soliton covers
a distance $\delta x=V\delta t+(n_{\mbox{\tiny{imp}}}\delta x)\, \Delta$.
Combining
this relation with Eq.~(\ref{e7}) one obtains a differential equation allowing
to determine $v=V/c$ as a function of $t$:
\begin{equation}\label{e11}
\frac{dv}{dt}=\frac{c}{4\,x_0}
\;\frac{F^+(v)+F^-(v)}{\sqrt{1-v^2}}
\;
\frac{1}{1-n_{\mbox{\tiny{imp}}}\,\Delta} \; .
\end{equation}
In the limit $v\to 1$, this equation
admits the analytical solution
\begin{equation}\label{e12}
t=\frac{15\,x_0}{c}\left\{G\left(v,\frac{n_{\mbox{\tiny{imp}}}\xi^2}{2b}\right)
-G\left(\frac{V_{\mbox{\tiny{init}}}}{c},
\frac{n_{\mbox{\tiny{imp}}}\xi^2}{2b}\right)
\right\} \; ,
\end{equation}
\noindent where 
\begin{eqnarray}\label{e13}
G(v,\alpha)& = & - \frac{\alpha}{v}+\frac{1+3\alpha}{4}
\ln\left(\frac{1+v}{1-v}\right)+\frac{1+\alpha}{2}\frac{v}{1-v^2} \nonumber \\ 
& \APPROX{v\to 1} &
\frac{1+\alpha}{4(1-v)}\; .
\end{eqnarray}
We compared this approximate result with the numerical
solution of (\ref{e11}) where $\Delta$ was evaluated through (\ref{e10}), 
and found that the accuracy of (\ref{e12}) is always
very good, even for initial velocities not close to $c$ (as was also the case
for the approximate expression (\ref{e8})).

The decay time $\tau$ of the soliton is the time at which $v=
V_{\mbox{\tiny{crit}}}/c \simeq 1-\frac{1}{2}(\xi/2a_1)^2$:
\begin{equation}\label{e14}
\tau=\frac{L}{c}\left(1+\frac{n_{\mbox{\tiny{imp}}}\xi^2}{2b}\right)
\; .
\end{equation}
In this expression -- as in (\ref{e9}) -- we neglected a corrective term
depending of the initial velocity, smaller by a factor $(\xi/a_1)^2$ than the
leading term. $\tau$ is proportional to $L/c$, with a slight
modification due to the shift induced at each scattering \cite{modif}:
repulsive obstacles ($b>0$) lead to an increased decay time since the soliton
covers the distance $L$ slightly more slowly that in the case of
attractive obstacles.

Formula (\ref{e14}) can be given a simple physical interpretation (in a less
rigorous setting) in the framework of the ``effective potential
approximation'' \cite{Bil05}. In this approximation, solitons are considered
as classical particles of mass $2m$ evolving in a potential
$U_{\mbox{\tiny{eff}}}$. One thus has $<\!\!m\,\dot{x}^2 +
U_{\mbox{\tiny{eff}}}(x) \!\!>= <\!\! m\,V^2(x) \!\!>$. The mean value of
$U_{\mbox{\tiny{eff}}}$ is the same as the one of $U$ \cite{mv} and from
Figure 2, one sees that at leading order it is sensible to approximate
$<\!\! m\,V^2(x) \!\!>$ by $m c^2=\mu$. One thus obtains $<\!\! \dot{x}^2
\!\!>\simeq c^2(1-<\!\! U(x)\!\!>/\mu)$. Finally, $\tau$ can be evaluated
through the formula
\begin{equation}\label{e15}
\tau=\frac{L}{<\!\! \dot{x}\!\!>}
\simeq\frac{L}{<\!\! \dot{x}^2\!\!>^{1/2}}\simeq
\frac{L}{c}\left(1+\frac{<\!\! U(x)\!\!>}{2\, \mu}\right) \; ,
\end{equation}
which is identical to (\ref{e14}). Since formulas (\ref{e9}) and (\ref{e15})
depend only on simple characteristics of the random potential (the average and
the two points correlation function), we expect them to be of very general
validity, poorly affected by the specific potential present in the disordered
region.

A final point to clarify is the effect of the random potential on the
occurrence of superfluidity and Bose-Einstein condensation, i.e., is Eq.
(\ref{e1}) truly applicable~? In the strong disorder limit, a quantum phase
transition occurs at $T=0$ leading to a (non superfluid) Bose glass phase
\cite{boseglass} where the description of the system with a single order
parameter $\psi(x,t)$ is inappropriate. However, in the case we consider here
of an atomic vapor described as a weakly interacting Bose gas, it has been
shown that a small amount of disorder does not drastically alter the
properties of the system, but merely decreases the condensate and the
superfluid fraction \cite{weakdis}. More precisely, based on the evaluations
presented in Ref.~\cite{Ast04} one can show that this effect is
negligible provided $n_{\mbox{\tiny{imp}}}\xi^3/b^2\ll 1$, which is the case
in the present study.

In conclusion, we have presented a description of the motion of a dark soliton
in a disordered region. The soliton radiates energy when it encounters an
obstacle. The repulsion between the particles has important consequences on
the propagation of the dark soliton, whose salient features are all at
variance with the one expected in the case of a linear wave packet or of a
bright soliton: (i) the soliton is {\it accelerated} to the velocity of sound
and disappears, (ii) its decay is algebraic, and (iii) the characteristic
decay length and decay time are independent of the initial velocity of the
soliton.

These results are generic and apply to many different fields (among which,
optics in nonlinear fibers with positive group velocity dispersion) but the
most promissing experimental relizations seem to be achievable for a Bose
condensed atomic vapor, either in a corrugated magnetic guide over a microchip
\cite{corrug}, or in an elongated trap in presence of an optical speckle
pattern \cite{Lye04}.

We acknowledge support form CNRS and Minist\`ere de la Recherche (Grant ACI
Nanoscience 201). Laboratoire de Physique Th\'eorique et Mod\`eles
Statistiques is Unit\'e Mixte de Recherche de l'Universit\'e Paris XI et du
CNRS, UMR 8626.


\begin{thebibliography}{99}
\bibitem{modif} The modification is small because
	$n_{\mbox{\tiny{imp}}}\xi^2/b \ll 1$.
\bibitem{mv} This is because $U_{\mbox{\tiny{eff}}}$ is the convolution of $U$
with a normalized function; see \cite{Bil05}.
\end{thebibliography}
\end{document}